\documentstyle[preprint,aps,eqsecnum]{revtex}

\tighten
\begin{document}
\draft

\title{
Boson-fermion mapping of collective fermion-pair algebras
}

\author  {  P. Navr\'atil$^{1,2}$, H.B. Geyer$^1$, and
            J. Dobaczewski$^{1,3}$
         }
\address { $^1$Institute of Theoretical Physics, University of
           Stellenbosch, 7599 Stellenbosch, South Africa\\
           $^2$Institute of Nuclear Physics, Czech Academy of
           Sciences, 250 68 \v{R}e\v{z} near Prague, Czech Republic\\
            $^3$Institute of Theoretical Physics, Warsaw University,
                       ul. Ho\.za 69, PL--00-681 Warsaw, Poland
        }

\maketitle

\begin{abstract}

We construct finite Dyson boson-fermion mappings of general collective
algebras extended by single-fermion operators.  A key element in the
construction is the implementation of a similarity transformation
which transforms boson-fermion images obtained directly from the
supercoherent state method. In addition to the general construction,
we give detailed applications to SO(2$N$), SU($\ell$+1), SO(5), and
SO(8) algebras.

\end{abstract}

\bigskip
\bigskip
\bigskip



\narrowtext

\section{Introduction}
\label{sec1}

The systematic construction of boson mappings for algebras defined by
collective bifermion operators, or more generally the construction of
boson realizations of Lie algebras, has by now been achieved in a
rather complete fashion for which a comprehensive literature exists.
(See Refs.\cite{Dob81-2,KM91,DDG+94} and references therein.)  As
shown by Dobaczewski\cite{Dob81-2}, one of the most
transparent ways to achieve this construction exploits generalized
coherent states\cite{Per86}, a formalism which has also recently been
shown\cite{DSG93,NGD94} to be readily amenable to the
construction of boson-fermion mappings from supercoherent states.

We refer to Ref.\cite{DSG93} for a background discussion and
motivation concerning the construction of boson-fermion mappings of
fermion systems in a many-body context.  Here it suffices to recall
that these methods aim at mapping (collective) fermion states with an
even number of particles onto boson states, while those with an odd
number of particles are mapped onto boson-fermion states.  In this way
odd boson-fermion states can be constructed in which collectivity and
the Pauli exclusion principle are consistently included in the same
theoretical framework.

In Ref.\cite{DSG93} it was possible to establish a general
prescription for constructing such mappings and in particular it was
shown that similarity transformations can be applied to modify
properties of boson-fermion images.  However, one important feature of
these images was missing, namely, the odd states were represented by
complicated many-fermion constructs instead of one-fermion states.
Here we report on the construction of a mapping which regains this
desired property and therefore corresponds to what has been hitherto
proposed in most phenomenological models for odd fermion states.  Some
aspects of our construction have already been implemented in
Ref.\cite{NGD94} where we discuss dynamical supersymmetry in
some specific fermion models.

It is thus the main objective of this paper to present the general
construction mentioned above and to give some examples.  The paper is
organized as follows.  In Sec.{\ }\ref{sec2} we begin by recalling some
basic definitions and notation and then we proceed by constructing the
similarity transformation which is the key element of the present
approach.  We then discuss general properties of this similarity
transformation when applied to the single-fermion images.
Section \ref{sec35} is devoted to examples concerning the
SO(2$N$), SU($\ell$+1), SO(5), and SO(8) algebras, respectively, and
conclusions are presented in Sec.{\ }\ref{sec7}.

\section{Dyson boson-fermion mapping of the collective algebra}
\label{sec2}

We consider the
boson-fermion
mapping of a collective algebra defined by the collective fermion-pair
creation operators
   \begin{equation}\label{e4}
   A^i = {\case{1}{2}}\chi^i_{\mu\nu} a^{\mu} a^{\nu} ,
   \end{equation}
labeled by the collective index $i$=$1,\ldots,M$.
Together with the
corresponding collective fermion-pair annihilation
operators, $A_i$=$(A^i)^+$, all linearly independent
commutators $\left[A_i,A^j\right]$, and the single-fermion
operators, $a^{\nu}$ and $a_{\nu}$, they are assumed to form a
closed collective superalgebra. (Here $\mu,\nu$=$1,\ldots,N$ and
$M\le N(N-1)/2$.) These closure conditions read
   \begin{equation}\begin{array}{rcl}\label{e5x}
   \left[\left[A_i,A^j\right],A_k\right] &=& c^{jl}_{ik}A_l
                                           , \\
   \left[A^i,a_{\nu}\right] &=& \chi^i_{\mu\nu} a^{\mu}
                                           , \\
   \left[A^i,a^{\nu} \right] &=& 0
                                           , \\
   \left\{a^{\mu},a_{\nu}\right\} &=& \delta^{\mu}_{\nu}
                                           , \\
   \left\{a^{\mu},a^{\nu} \right\} &=& 0
                                           , \\
   \end{array}\end{equation}
where $c^{jl}_{ik}$ are structure constants and an implicit summation
over repeated indices is assumed.  With the single-fermion operators
anti-commuting to the identity, the odd and even parts of this superalgebra
are composed of single-fermion and bifermion operators, respectively.

Following Ref.{\ }\cite{DGH91}, we assume that the
collective pairs are orthogonal and normalized to a common
number $g$, i.e.,
 \begin{equation}\label{e60}
   \langle 0| A_i A^j|0\rangle \equiv
        {\case{1}{2}}\chi_i^{\mu\nu}\chi^j_{\mu\nu}
        = g\delta^j_i ,
 \end{equation}
$\big(\chi_i^{\mu\nu}$=$(\chi^i_{\mu\nu})^{\displaystyle\ast}\big)$
which gives the symmetry properties of structure constants
 \begin{equation}\label{e62}
   c^{jl}_{ik} = c^{lj}_{ik} =  c^{jl}_{ki}
                             = (c^{ik}_{jl})^{\displaystyle\ast} .
 \end{equation}

A Dyson type boson-fermion mapping of this algebra was derived in
Ref.\cite{DSG93}
using a collective Usui operator
   \begin{equation}\label{e125}
   U = \langle{0}|\exp\left(B^i A_i + \alpha^{\mu} a_{\mu}\right)|0)
   \end{equation}
suggested by the supercoherent state method. This operator
transforms collective even-fermion states, as well as
collective states with additional individual fermions, into an
ideal space composed of collective bosons,
$B^i$=$B_i^{\mbox{\footnotesize{\dag}}}$,
$[B_j,B^i]$=$\delta^i_j$, and of ideal
fermions $\alpha^{\mu}$.

The mapping of operators
${\cal O}$$\longleftrightarrow$$\hat O$
can be obtained from the equation
${\cal O} U$=$U\hat O$, which gives the following mapping of the
collective superalgebra (\ref{e5x}):
   \begin{mathletters}\label{preimages}
   \begin{eqnarray}
   A^j               &\longleftrightarrow&
                         gB^j - {\case{1}{2}}
                     c^{jl}_{ik}B^i B^k B_l
     - \chi^j_{\mu\rho}\chi_i^{\nu\rho}
                     B^i\alpha^{\mu}\alpha_{\nu}
     + {\cal A}^j \nonumber \\
   &&=[\Lambda,B^j] + {\cal A}^j , \label{e126c} \\
   A_j  &\longleftrightarrow&
   B_j      , \label{e126a} \\
   \big[A_i,A^j\big] &\longleftrightarrow&
                         g\delta^j_i - c^{jl}_{ik}B^kB_l
                       - \chi^j_{\mu\rho}\chi_i^{\nu\rho}
                     \alpha^{\mu}\alpha_{\nu}
                                              , \label{e126b} \\
   a^{\nu}           &\longleftrightarrow&
                         \alpha^{\nu}
                   + \chi_i^{\nu\rho} B^i\alpha_{\rho}
                                              , \label{e126e} \\
   a_{\nu}           &\longleftrightarrow&
                         \alpha_{\nu}
                                              . \label{e126d}
   \end{eqnarray}\end{mathletters}
where we introduced collective ideal fermion pairs,
   \begin{equation}\label{e127}
   {\cal A}^j = {\case{1}{2}}
        \chi^j_{\mu\nu}\alpha^{\mu}\alpha^{\nu} ,
        \quad
        {\cal A}_j$=$({\cal A}^j)^{\mbox{\footnotesize{\dag}}}\; ,
   \end{equation}
and the operator $\Lambda$,
\begin{equation}\label{eq5a}
\Lambda = g B^l B_l -{\textstyle{1\over 4}} c_{m n}^{k l} B^m B^n B_l B_k
          - \chi^k_{\mu\rho}\chi^{\nu\rho}_l B^l B_k \alpha^\mu
           \alpha_\nu\; ,
\end{equation}
invariant with respect to the core subalgebra.

At this point it is important to recall that a key element in
the boson-fermion mapping formalism is the existence of a {\em
physical subspace} of the ideal space. This subspace is the one
whose states can be put into a one-to-one correspondence with the
original fermion space. In Dyson type mappings (as opposed to
e.g.{\ }Schwinger
type mappings\cite{SG88}) this one-to-one correspondence is
guaranteed by simply operating (repeatedly) with bifermion and single
fermion images onto the ideal space vacuum. At the same time the
nature of the physical states obtained from this construction will
clearly depend on the structure of the images and may not necessarily
correspond to some preconceived physically desirable structure. This
point was discussed in  Ref.\cite{DSG93} and is illustrated by the
images (\ref{preimages}).

Consider e.g.{\ }the image of the collective pair operator $A^i$,
Eq.{\ }(\ref{e126c}), which contains the corresponding ideal collective pair
operator ${\cal A}^i$.  Operating with this image onto the ideal space
vaccum $|0)$ gives $A^i|0\rangle\leftrightarrow ({\cal A}^i+gB^i)|0)$,
and the collective one-pair states are therefore not completely
bosonized as one would desire from a physical point of view. It
is necessary, therefore, to transform the mapping (\ref{preimages})
into a form which will suitably address this problem, namely
 result in a description where collective fermion pairs are
represented
by bosons only, while all other fermion degrees of freedom are simply
accommodated as ideal fermions.  Furthermore, the ideal states should
accommodate an arbitrary number of ideal fermions, representing
non-collective fermions, but of course still subject to reigning space
limitations.

An attempt was made in the Ref.\cite{DSG93} to find a transformation
leading to a mapping which reduces to the standard Dyson
mapping (with bosons only) if the odd degrees of freedom
are dropped.  The drawback of the transformation suggested was that
the mapping resulted in images of single-fermion operators
given as infinite series.  At the same time, the odd states were
mapped onto rather complicated boson-fermion states.

In Ref.\cite{NGD94} we presented a similarity transformation which
resolves this problem, i.e., gives even collective ideal states
which are completely bosonized and at the same time yields the
odd states described by single ideal fermions.  In that paper we
proved by induction that a specific similarity transformation meets
these requirements.  Here we proceed differently, by showing
how the same transformation can be derived
from properties of some hamiltonian-like operators.

\subsection{Construction of the transformation}

We rewrite the right-hand sides of expressions
(\ref{e126c}) and (\ref{e126b})
in a short-hand notation as
\begin{mathletters}\begin{eqnarray}
A^j   &\longleftrightarrow& R^j + {\cal A}^j , \label{eq1} \\
A_{j} &\longleftrightarrow& B_j ,   \\
\left[A_i,A^j\right] &\longleftrightarrow&
      \left[B_i,R^j + {\cal A}^j\right] . \label{eq1c}
\end{eqnarray}\end{mathletters}%
Firstly, we observe that because
$\left[B_i,{\cal A}^j\right]$=0, we may
simply drop the term ${\cal A}^j$
in Eqs.{\ }(\ref{eq1}) and (\ref{eq1c}),
and the commutation relations of the
collective algebra will still be
satisfied, i.e., the operator $R^j$ alone when commuted with $B_j$
gives the right-hand side of Eq.{\ }(\ref{e126b}).
We may, therefore, envisage
a similarity transformation ($X$, say)
that will achieve just this modification,
namely,
\begin{mathletters}\label{eq22}\begin{eqnarray}
X^{-1} \left(R^j + {\cal A}^j\right) X &=& R^j , \label{eq22a} \\
X^{-1} B_j                           X &=& B_j , \label{eq22b}
\end{eqnarray}\end{mathletters}
and which could then be applied to the right-hand sides of
expressions
(\ref{e126e}) and (\ref{e126d})
to find the corresponding single-fermion images.

To evaluate $X$ we first multiply
expression (\ref{eq1}) by $B_j$ and sum over $j$. This gives
the operator
\begin{equation}\label{eq2a}
H \equiv R^j B_j + {\cal A}^j B_j ,
\end{equation}
where the first term on the right-hand-side
conserves the number of bosons
and
ideal fermions separately, while the second term decreases
the number
of bosons by one and increases the number of ideal fermions by two.
It is clear that this operator
has an upper (or lower) triangular structure in the basis
characterized
by the numbers of ideal bosons (or fermions). Consequently, it has the
same
spectrum as the operator
\begin{equation}\label{eq2b}
H_0 \equiv R^j B_j
\end{equation}
and, therefore, $H$ and $H_0$ are
related by a similarity transformation of the type introduced by Geyer
\cite{Gey86} %
\begin{equation}\label{eq3}
 X = \sum_{k=0}^{\infty} X_k
   = \sum_{k=0}^{\infty} (\frac{1}{\widehat{H}_{0} - H_{0}}
{\cal A}^j B_j)^{k}\:
{\textstyle \raisebox{-1ex}{$\widehat{}$}}\qquad,
\end{equation}
i.e.,
\begin{equation}\label{eq3x}
X^{-1} H X = H_0 .
\end{equation}
Here the lone hat ``$\:\:$
\raisebox{-.8ex}{$\widehat{}$}$\:\:\:$"
is read together with a positional operator\cite{KV87} and determines
at which position the hatted operator,
$\widehat{H}_{0}$ in this case, is
evaluated.

The denominator in expression (\ref{eq3}) can be written in a
convenient form much more directly linked to the structure of the
ideal fermion (core) algebra. To show this we first identify
\begin{eqnarray}
C_{\rm B} &=& g B^l B_l - \case{1}{2} c_{m n}^{k l} B^m B^n B_k B_l
                                           \nonumber \\
    &=& g B^l B_l - \case{1}{2} c_{m n}^{k l} B^m B_k B^n B_l
                  + \case{1}{2} c_{m l}^{k l} B^m B_k
                                                 \label{eq3a}
\end{eqnarray}
as the invariant operator of the boson core algebra, and
\begin{equation}\label{eq3b}
C_{\rm F} = {\cal A}^l {\cal A}_l
\end{equation}
as the invariant operator of the ideal   fermion core
algebra. The invariant
operator of the boson-fermion core algebra may be conveniently
expressed if we perform an {\em auxiliary bosonization} of the
ideal fermion pair algebra by using auxiliary bosons $b^i$ and
$b_j$ which commute
with the other boson operators $B$,
   \begin{mathletters}\label{eqsecbos}\begin{eqnarray}
   {\cal A}^j               &\longleftrightarrow&
                         gb^j - {\case{1}{2}}
                     c^{jl}_{ik}b^i b^k b_l     , \label{e128c} \\
   {\cal A}_j  &\longleftrightarrow&
                                        b_j     , \label{e128a} \\
   \big[{\cal A}_i,{\cal A}^j\big] &\longleftrightarrow&
                         g\delta^j_i - c^{jl}_{ik}b^kb_l
                                                . \label{e128b}
   \end{eqnarray}\end{mathletters}%
This greatly facilitates
the construction of
the boson-fermion invariant operator, since the similar structures of
$C_{\rm B}$ and $C_{\rm F}$ (when the latter is expressed in terms of the
auxiliary bosons)
allows one to write in analogy of expression
(\ref{eq3a})
\begin{eqnarray}
C_{\rm BF} &=& g (B^l B_l+b^l b_l) \nonumber \\
&-& \case{1}{2} c_{m n}^{k l} (B^m B_k +b^m b_k)(B^n B_l+b^n b_l)
                              \label{eq3c}    \nonumber \\
&+& \case{1}{2} c_{m l}^{k l} (B^m B_k +b^m b_k)\; .
\end{eqnarray}
This can be further rewritten as
\begin{eqnarray}\label{eq3d}
C_{\rm BF} &=& C_{\rm F} + B^l [{\cal A}_l,{\cal A}^k] B_k
- \case{1}{2} c_{m n}^{k l} B^m B^n B_k B_l  \nonumber \\
           &=& C_{\rm F} + H_0 ,
\end{eqnarray}
with $C_{\rm F}$ now again given by expression (\ref{eq3b}). It follows that
$H_0$=$C_{\rm BF}$$-$$C_{\rm F}$, and since
\begin{equation}\label{eqn3e}
\left[ C_{\rm BF} , {\cal A}^j B_j \right] = 0
\end{equation}
the desired result is
\begin{equation}\label{eq3e}
\widehat{H}_0 - H_{0} = C_{\rm F} - \widehat{C}_{\rm F} ,
\end{equation}
which explicitly demonstrates that the denominator in (\ref{eq3})
depends only on the invariant operators of the
ideal fermion core algebra, i.e.,
\begin{equation}\label{eq3y}
 X = \sum_{k=0}^{\infty} X_k
   = \sum_{k=0}^{\infty} (\frac{1}{C_{\rm F} - \widehat{C}_{\rm F}}
{\cal A}^j B_j)^{k}\:
{\textstyle \raisebox{-1ex}{$\widehat{}$}}\qquad.
\end{equation}

We were not able to derive a
similar closed form of the inverse transformation
$X^{-1}$, but in fact this is not necessary
provided we know how to commute boson-fermion operators with $X$.
Indeed, ${\cal O}'$ is the similarity transform of the operator ${\cal
O}$ if it satisfies the equation ${\cal O}X$=$X{\cal O}'$ where only
the operator $X$ appears. By inspection of the structure of $X$,
however, it is possible to write down the lowest order terms in an
expansion for $X^{-1}$,
\begin{eqnarray}\label{eq3z}
 X^{-1} = 1 &-& \frac{1}{C_{\rm F} - \widehat{C}_{\rm F}} {\cal A}^j B_j
    \: {\textstyle \raisebox{-1ex}{$\widehat{}$}} \:\: \nonumber \\
        &+&\left( \frac{1}{C_{\rm F} - \widehat{C}_{\rm F}} {\cal A}^j B_j
    \: {\textstyle \raisebox{-1ex}{$\widehat{}$}} \:\:\right)
   \left(\frac{1}{C_{\rm F} - \widehat{C}_{\rm F}} {\cal A}^i B_i
    \: {\textstyle \raisebox{-1ex}{$\widehat{}$}} \:\:\right)
 \nonumber \\
 &-&           \frac{1}{C_{\rm F} - \widehat{C}_{\rm F}} {\cal A}^j B_j \:
   \frac{1}{C_{\rm F} - \widehat{C}_{\rm F}} {\cal A}^i B_i
    \: {\textstyle \raisebox{-1ex}{$\widehat{}$}}
\;\;\ldots,
\end{eqnarray}
where brackets demarcate where the
hatted operators are to be evaluated in
a term with more than one of these hat indicators,
``$\:\:{\textstyle \raisebox{-1ex}{$\widehat{}$}} \:\:$".

It should be noted that $X$ and $X^{-1}$ have an identical
triangular structure, namely, they contain terms which decrease
(increase) boson (fermion) numbers by zero, one, two, etc.{\ }
(zero, two, four, etc.).

After deriving the expression for $X$ we may now verify
Eqs.{\ }(\ref{eq22}). A proof of these identities by
induction has been
given in Ref.{\ }\cite{NGD94} and will not be repeated here.
While it is clear that the transformation law
(\ref{eq3x}) for $H$  is a consequence of the  transformation
laws (\ref{eq22}) for $R^j$+${\cal A}^j$ and $B_j$,
it was, however, only
by considering the hamiltonian $H$ that we were actually
able to derive the similarity transformation in the form of
Eqs.{\ }(\ref{eq3}) and (\ref{eq3y}).

\subsection{Single-fermion images}
\label{single}

To find the Dyson boson-fermion images of single-fermion operators
we apply the transformation (\ref{eq3y}) to (\ref{e126e}) and (\ref{e126d}).
{}From Eqs.{\ }(\ref{eq3y}) and (\ref{eq3z}) it follows that the
transformed annihilation operator
\widetext
\begin{eqnarray}
& &X^{-1} \alpha_\nu X = \alpha_\nu
- \frac{1}{C_{\rm F} - \widehat{C}_{\rm F}}{\cal A}^l B_l\:
{\textstyle \raisebox{-1ex}{$\widehat{}$}}\: \: \alpha_\nu
+ \alpha_\nu \frac{1}{C_{\rm F} - \widehat{C}_{\rm F}}{\cal A}^l B_l\:
{\textstyle \raisebox{-1ex}{$\widehat{}$}}\qquad
                                        \nonumber \\
&-& \frac{1}{C_{\rm F} - \widehat{C}_{\rm F}}{\cal A}^l B_l \:
    \frac{1}{C_{\rm F} - \widehat{C}_{\rm F}}{\cal A}^m B_m \:
\:{\textstyle \raisebox{-1ex}{$\widehat{}$}}\:\: \alpha_\nu
+  \alpha_\nu \frac{1}{C_{\rm F} - \widehat{C}_{\rm F}}{\cal A}^l B_l \:
    \frac{1}{C_{\rm F} - \widehat{C}_{\rm F}}{\cal A}^m B_m \:
\:{\textstyle \raisebox{-1ex}{$\widehat{}$}}\:\:
                                      \nonumber \\
&-& \left(\frac{1}{C_{\rm F} - \widehat{C}_{\rm F}}{\cal A}^l B_l \:
    \:{\textstyle \raisebox{-1ex}{$\widehat{}$}}\:\: \right)
    \alpha_\nu
    \left(\frac{1}{C_{\rm F} - \widehat{C}_{\rm F}}{\cal A}^m B_m \:
\:{\textstyle \raisebox{-1ex}{$\widehat{}$}}\:\:\right)
                                       \nonumber \\
&+&
   \left( \frac{1}{C_{\rm F} - \widehat{C}_{\rm F}}{\cal A}^l B_l \:
    \:{\textstyle \raisebox{-1ex}{$\widehat{}$}}\:\:\right)
   \left( \frac{1}{C_{\rm F} - \widehat{C}_{\rm F}}{\cal A}^m B_m \:
   \:{\textstyle \raisebox{-1ex}{$\widehat{}$}}\:\: \right)
    \alpha_\nu
   + \ldots \label{eqx4}
\end{eqnarray}
decreases the number of ideal fermions by one and increases the number
of ideal fermions by 1, 3, 5 etc. The same is true for
the transformed creation operator
\begin{eqnarray}
& &X^{-1} (\alpha^\nu + \chi_n^{\nu\rho}B^n \alpha_\rho) X =
\alpha^\nu + \chi_n^{\nu\rho}B^n \alpha_\rho
- \frac{1}{C_{\rm F} - \widehat{C}_{\rm F}}{\cal A}^l B_l\:
\:{\textstyle \raisebox{-1ex}{$\widehat{}$}}\:\:
\chi_n^{\nu\rho}B^n \alpha_\rho
                                        \nonumber \\
&+& \chi_n^{\nu\rho}B^n \alpha_\rho \frac{1}{C_{\rm F} - \widehat{C}_{\rm F}}
{\cal A}^l B_l\:
\:{\textstyle \raisebox{-1ex}{$\widehat{}$}}\:\:
- \frac{1}{C_{\rm F} - \widehat{C}_{\rm F}}{\cal A}^l B_l \:
\:{\textstyle \raisebox{-1ex}{$\widehat{}$}}\:\: \alpha^\nu
+ \alpha^\nu \frac{1}{C_{\rm F} - \widehat{C}_{\rm F}}{\cal A}^l B_l \:
\:{\textstyle \raisebox{-1ex}{$\widehat{}$}}\:
                                         \nonumber \\
&-&\frac{1}{C_{\rm F} - \widehat{C}_{\rm F}}{\cal A}^l B_l \:
    \frac{1}{C_{\rm F} - \widehat{C}_{\rm F}}{\cal A}^m B_m \:
\:{\textstyle \raisebox{-1ex}{$\widehat{}$}}\:\:
\chi_n^{\nu\rho}B^n \alpha_\rho
   +  \chi_n^{\nu\rho}B^n \alpha_\rho
    \frac{1}{C_{\rm F} - \widehat{C}_{\rm F}}{\cal A}^l B_l \:
    \frac{1}{C_{\rm F} - \widehat{C}_{\rm F}}{\cal A}^m B_m \:
\:{\textstyle \raisebox{-1ex}{$\widehat{}$}}\:\:
                                      \nonumber \\
&-&\left( \frac{1}{C_{\rm F} - \widehat{C}_{\rm F}}{\cal A}^l B_l \:
    \:{\textstyle \raisebox{-1ex}{$\widehat{}$}}\:\:\right)
    \chi_n^{\nu\rho}B^n \alpha_\rho
    \left(\frac{1}{C_{\rm F} - \widehat{C}_{\rm F}}{\cal A}^m B_m \:
    \:{\textstyle \raisebox{-1ex}{$\widehat{}$}}\:\:\right)
                                       \nonumber \\
&+&
   \left(\frac{1}{C_{\rm F} - \widehat{C}_{\rm F}}{\cal A}^l B_l \:
    \:{\textstyle \raisebox{-1ex}{$\widehat{}$}}\:\:\right)
    \left(\frac{1}{C_{\rm F} - \widehat{C}_{\rm F}}{\cal A}^m B_m \:
     \:{\textstyle \raisebox{-1ex}{$\widehat{}$}}\:\:\right)
    \chi_n^{\nu\rho}B^n \alpha_\rho
+ \ldots \quad.\label{eqx5}
\end{eqnarray}\narrowtext\noindent%
In Eqs.{\ }(\ref{eqx4}) and (\ref{eqx5})
we show all terms changing the number
of ideal fermions by one and increasing this number by three.

The question arises whether the terms
in the single-fermion operator images
which increase the ideal
fermion number by more than one
can contribute to matrix elements in the full ideal
boson-fermion space. Although we were not able to
settle this question on the operator level, we will demonstrate
that the answer is negative for at least a wide class of states
characterized by the condition
\begin{equation}\label{eqn20}
{\cal A}_l |\psi) = 0
\end{equation}
for all those $l$ which refer to the collective pairs
singled out
for mapping onto bosons as in (\ref{e126c}).

On the one hand this is a
physically relevant condition, since the mapping should be designed to
eliminate collective ideal fermion pairs in favor of bosons. At the
same time condition (\ref{eqn20}) does not
limit the considerations to the physical subspace of the
ideal boson-fermion space, as is often the case with ideal space
relations which are the counterparts of operator identities in the
original space. (See also an explicit example in
Sec.{\ }\ref{sec5a}.)
We note that the condition (\ref{eqn20}) is
automatically fulfilled for the most practically interesting
cases of $n_{\rm F}=0$
and $n_{\rm F}=1$. For ideal space states with two ideal fermions, the
condition implies, as anticipated above, that the ideal
fermions should form pairs orthogonal to all the collective pairs
mapped onto bosons.

Condition (\ref{eqn20}) and its consequences for the single-fermion
images are explicitly discussed in Appendix \ref{appA}. Here we only
collect the final expressions which
define the Dyson boson-fermion
mapping of the general collective algebra characterized by
(anti)commutation relations (\ref{e5x}):
\widetext
\begin{mathletters}\begin{eqnarray}
   A^j               &\longleftrightarrow&
                         gB^j - {\case{1}{2}}
                     c^{jl}_{ik}B^i B^k B_l
                     - \chi^j_{\mu\rho}\chi_i^{\nu\rho}
                     B^i\alpha^{\mu}\alpha_{\nu}
                                              , \label{s126c} \\
   A_j            &\longleftrightarrow&
                         B_j                  , \label{s126a} \\
   \big[A_i,A^j\big] &\longleftrightarrow&
                         g\delta^j_i - c^{jl}_{ik}B^kB_l
                       - \chi^j_{\mu\rho}\chi_i^{\nu\rho}
                     \alpha^{\mu}\alpha_{\nu}
                                              , \label{s126b} \\
   a^{\nu}           &\longleftrightarrow&
                         \alpha^{\nu}
                   + \chi_i^{\nu\rho} B^i\alpha_{\rho}
           - \frac{1}{C_{\rm F} - \widehat{C}_{\rm F}}{\cal A}^l B_l\:
\:{\textstyle \raisebox{-1ex}{$\widehat{}$}}\:\:
\chi_n^{\nu\rho}B^n \alpha_\rho                 \nonumber  \\
& & + \chi_n^{\nu\rho}B^n \alpha_\rho \frac{1}{C_{\rm F} - \widehat{C}_{\rm F}}
{\cal A}^l B_l\:
\:{\textstyle \raisebox{-1ex}{$\widehat{}$}}\:\:\ldots
                                              , \label{s126e} \\
   a_{\nu}           &\longleftrightarrow&
                         \alpha_{\nu}
- \frac{1}{C_{\rm F} - \widehat{C}_{\rm F}}{\cal A}^l B_l\:
{\textstyle \raisebox{-1ex}{$\widehat{}$}}\: \: \alpha_\nu
 + \alpha_\nu \frac{1}{C_{\rm F} - \widehat{C}_{\rm F}}{\cal A}^l B_l\:
{\textstyle \raisebox{-1ex}{$\widehat{}$}}\:\:\ldots
                                              , \label{s126d}
\end{eqnarray}\end{mathletters}\narrowtext\noindent%
where the dots $\ldots$
refer in general to higher-order terms
increasing the number of ideal fermions by 3, 5, etc.
These terms cancel when
acting on a wide
class of states characterized by condition (\ref{eqn20}).

\subsection{Physical subspace}

We now discuss the structure of the physical states and show how
the condition (\ref{eqn20}) is satisfied for them.  Simultaneously,
{it becomes clear that the higher-order terms in the
single-fermion operator image cannot contribute to the physical states.
First we note that the similarity transformation (\ref{eq3y}) does
not
change any state in which there are no bosons $B^j$, and therefore the
single-fermion states are mapped onto single ideal fermion states,
\begin{equation}\label{eqn6}
a^{\nu}|0\rangle\leftrightarrow\alpha^{\nu}|0) .
\end{equation}
Since the images of
pair creation operators (\ref{s126c})
do not change the ideal fermion
number, collective odd states $ |\Psi_{\text{odd}}\rangle$ will be
mapped onto ideal states with one ideal fermion only.  Especially from
a physical point of view, this result is a clear improvement over the
solutions found in Ref.\ \cite{DSG93}, where collective odd
states were mapped onto ideal states with many-fermion components.

Repeated application of (\ref{s126e}) shows that the two-fermion
states are mapped as
\begin{equation}\label{eqn7}
      a^{\mu}a^{\nu}|0\rangle\longleftrightarrow
      \bigl(\alpha^{\mu}\alpha^{\nu}+\chi^{\mu\nu}_iB^i
     -\frac{1}{C_{\rm F}}\chi^{\mu\nu}_i{\cal{A}}^i\bigr)|0) ,
\end{equation}
and in general contain the non-collective pair of ideal fermions
$\alpha^{\mu}\alpha^{\nu}|0)$. However, when the collective pair
$A^j$ is formed by summing the pairs $a^{\mu}a^{\nu}$
with collective amplitudes ${\textstyle{1\over 2}}\chi^j_{\mu\nu}$, the ideal
non-collective pairs above recombine and since
[see Eq.{\ }(\ref{e60})]
$C_{\rm F}{\cal{A}}^j|0)$=$g{\cal{A}}^j|0)$
the first and last terms above cancel. Only the boson state
$gB^j|0)$ therefore remains as the image of a collective fermion pair
state.  Again, since the images (\ref{s126c}) conserve the
ideal fermion number, the same recombination mechanism is also valid
for any even state.  It is also clear that ${\cal A}_l$ acting
on the right-hand-side of (\ref{eqn7}) gives zero,
in accordance with the
condition (\ref{eqn20}).

Consider now the images of three-fermion states. From
(\ref{eqx5}) and (\ref{eqn7}) we observe that
the terms increasing the ideal fermion number by three could
also contribute to a three-fermion state
constructed by an application of (\ref{eqx5}) on
(\ref{eqn7}).
These terms cancel as follows from the proof in the preceding section,
because here these terms would act on a state with no ideal fermion.
The three-fermion state therefore contains
contributions of only those terms shown explicitly
in (\ref{s126e}) and can be written as
\widetext
\begin{eqnarray}
a^{\tau}a^{\mu}a^{\nu}|0\rangle&\longleftrightarrow&
      \bigl(\alpha^{\tau}\alpha^{\mu}\alpha^{\nu}
     +\alpha^{\tau}\chi_{\rho\tau}^lB^i
     +\alpha^{\nu}\chi^{\tau\mu}_iB^i
     +\alpha^{\mu}\chi^{\nu\tau}_iB^i
     \nonumber \\
     &-&\alpha^{\tau}\frac{1}{C_{\rm F}}\chi^{\mu\nu}_i{\cal{A}}^i
     -\frac{1}{C_{\rm F}}\chi^{\tau\mu}_i{\cal{A}}^i \alpha^{\nu}
     -\frac{1}{C_{\rm F}}\chi^{\nu\tau}_i{\cal{A}}^i \alpha^{\mu}
     \nonumber \\
     &+&\frac{1}{g}\frac{1}{C_{\rm F}}\chi_{\rho\tau}^l\chi^{\nu\mu}_i
     \chi_{\rho\sigma}^i{\cal{A}}^l \alpha^{\sigma}
     \bigr)|0)  .     \label{eqn11}
\end{eqnarray}\narrowtext\noindent%
It is simple to verify that $\chi_{\mu\nu}^i a^{\tau}a^{\mu}a^{\nu}|0\rangle$
maps onto $R^i \alpha^\tau|0)$, but it is more involved to show that
the image of, say,
$\chi_{\tau\mu}^i a^{\tau}a^{\mu}a^{\nu}|0\rangle$, obtained by
explicitly combining the three indicated images, reduces  to $R^i
\alpha^\nu|0)$, as it should if the mapping is consistent.
This calculation can again be facilitated by
performing an
auxiliary bosonization of the ideal fermion operators using
(\ref{eqsecbosfer}).
Once more the higher-order terms do not contribute when acting
as in (\ref{eqn11}) as it fulfills
the condition
${\cal A}_l |\psi)=0$, which can best be recognized in the
auxiliary bosonized picture.

So far we have discussed states constructed from the single-fermion
operator (\ref{s126e}).  However, because the pair creation operator
image (\ref{s126c}) does not change the ideal fermion number and
commutes with (\ref{s126e}) (which can be checked explicitly) our
previous analysis is valid for any physical state obtained by the
application of (\ref{s126c}) on the states (\ref{eqn6}),
(\ref{eqn7}),
and (\ref{eqn11}).  In particular, the terms increasing the ideal
fermion number by three, five etc.{\ }do not contribute to these physical
states.

One property of the single-fermion images that has so far not
been discussed, concerns the question whether {\it anti-}commutators
among the original fermion operators $a^\mu$ and $a_\nu$ are preserved
on the ideal space by their images.  Common wisdom has so far held
that it is only on the physical subspace that the images can in fact
preserve these anti-commutators \cite{KM91,GH80a,GH83,Oku74}.  This
conclusion mostly follows from the particular construction of images
and the subsequent method of verifying these relations in the ideal
space, which typically leads to results of the type $\{(a_\nu)_{\rm
I}, (a^\mu)_{\rm I}\}P = \delta^\mu_\nu P$, where the subscript I
denotes a general image and $P$ the projection operator to the
physical subspace of the full ideal space.

One of the advantages of our construction through supercoherent states
is that it is easily verified that the images (\ref{e126e}) and
(\ref{e126d}) preserve anti-commutation relations on the {\it full}
ideal space. Subsequent images obtained from these original images
through similarity transformation will naturally retain this property,
as long as the transformation $X$, which defines the similarity
transform $\cal O'$ of operator $\cal O$ through
${\cal O}X=X{\cal O}'$, is non-singular, as is the case in our
applications. We return to this point in the following examples.

\section{Applications and examples}
\label{sec35}

The mapping derived in the previous section also covers the
non-collective algebra SO(2$N$).
As we discuss in Sec.{\ }\ref{sec3},
the similarity transformation may in this case be
expressed in a more compact form. Other examples,
pertaining to collective algebras, are presented in
Secs.{\ }\ref{sec4}--\ref{sec5}.

\subsection{SO(2$N$) mapping}
\label{sec3}

The Dyson boson-fermion mapping
obtained in Ref.\cite{DSG93} from the
supercoherent state method for SO(2$N$) is
\begin{mathletters}\begin{eqnarray}
   a^{\mu} a^{\nu} &\longleftrightarrow&
                       B^{\mu\nu}
                 - B^{\mu\rho} B^{\nu\theta} B_{\rho\theta} \nonumber \\
               &&- B^{\mu\rho}\alpha^{\nu}\alpha_{\rho}
                     + B^{\nu\rho}\alpha^{\mu}\alpha_{\rho}
                     + \alpha^{\mu}\alpha^{\nu} ,
                                \label{e110c} \\
   a_{\nu} a_{\mu} &\longleftrightarrow&
                       B_{\mu\nu}               ,
                                \label{e110a} \\
   a^{\mu} a_{\nu} &\longleftrightarrow&
                       B^{\mu\theta} B_{\nu\theta}
                 + \alpha^{\mu}\alpha_{\nu} ,
                                \label{e110b} \\
   a^{\nu}         &\longleftrightarrow&
                       \alpha^{\nu}
                 + B^{\nu\rho}\alpha_{\rho} ,
                                \label{e110e} \\
   a_{\nu}      &\longleftrightarrow&
                       \alpha_{\nu}             .
                                \label{e110d}
\end{eqnarray}\end{mathletters}%

The invariant operator of the ideal fermion core algebra depends in
this
case only on the number operator $n$=$\alpha^\rho \alpha_\rho$,
namely,
\begin{equation}
C_{\rm F} = \case{1}{2}\alpha^{\mu}\alpha^{\nu}\alpha_{\nu}\alpha_{\mu}
  = \case{1}{2}n(n-1).
\end{equation}
This simplifies the similarity transformation significantly.
In particular we have
\begin{equation}\label{eqso1}
C_{\rm F} - \widehat{C}_{\rm F} = \case{1}{2}(n-\hat{n})(n+\hat{n}-1)
,
\end{equation}
and the transformation (\ref{eq3y}) can be summed to the
following closed
form
\begin{equation}\label{eqso2}
 X = \frac{(2\hat{n}-1)!!}{(n+\hat{n}-1)!!} {\exp}\left[ \case{1}{2}
\alpha^\mu \alpha^\nu B_{\mu\nu}\right] \:
{\textstyle \raisebox{-1ex}{$\widehat{}$}}\qquad.
\end{equation}
Finally, the transformed Dyson boson-fermion mapping is
\widetext
\begin{mathletters}\label{eqsso2n}\begin{eqnarray}
a^{\mu} a^{\nu} &\longleftrightarrow&
                       B^{\mu\nu}
                 - B^{\mu\rho} B^{\nu\theta} B_{\rho\theta}
                 - B^{\mu\rho}\alpha^{\nu}\alpha_{\rho}
                     + B^{\nu\rho}\alpha^{\mu}\alpha_{\rho}
                      ,
                                \label{so110c} \\
a_{\nu} a_{\mu} &\longleftrightarrow&
                       B_{\mu\nu}               ,
                                \label{so110a} \\
a^{\mu} a_{\nu} &\longleftrightarrow&
                       B^{\mu\theta} B_{\nu\theta}
                 + \alpha^{\mu}\alpha_{\nu} ,
                                \label{so110b} \\
a^{\nu}   &\longleftrightarrow&
                       \alpha^{\nu}
                 + B^{\nu\rho}\alpha_{\rho}
     -\frac{1}{2n-3}\alpha^\nu n + \frac{1}{2n-1}\alpha^\tau
      B^{\nu\rho}B_{\rho\tau}
                 \nonumber \\ &&
    -\frac{1}{(2n-1)(2n-3)}\alpha^\sigma\alpha^\tau\alpha_\rho
                                B^{\nu\rho}B_{\sigma\tau} ,
                                \label{so110e} \\
   a_{\nu}      &\longleftrightarrow&
                       \alpha_{\nu}
-\frac{1}{(2n-1)(2n-3)}\alpha^\sigma\alpha^\tau \alpha_\nu B_{\sigma\tau}
+ \frac{1}{2n-1}\alpha^\tau B_{\nu\tau} .
                                \label{so110d}
\end{eqnarray}\end{mathletters}\narrowtext\noindent%
We observe that the single-fermion images are
finite and comprised of
terms changing the ideal fermion number by one only.
Moreover, the mapping is such that an attempt to create a fermion pair
using successive application of (\ref{so110e}) is equivalent
to the application of (\ref{so110c}).

It is not difficult to verify, as anticipated in the previous
section, that the images (\ref{so110e}) and (\ref{so110d}) preserve
anti-commutators on the full ideal space, i.e.\ as {\it operator}
identities. An efficient way to do this is first to form a product of
two single fermion images and then to {\it symmetrize} with respect to
the indices. Upon symmetrization some terms in the product will
immediately yield zero on their own, because of their
anti-symmetry, while others will conspire to yield either zero or
unity, depending on the anti-commutator being verified.

\subsection{SU($\ell$+1) mapping}
\label{sec4}

Let us suppose that the collective operators form an
$(\Omega+1)$-dimensional symmetric representation of the
unitary algebra SU($\ell$+1), i.e., one has $\ell$ collective
pairs $A_i$.
This model can be realized in an ($\ell$+1)$\times$$\Omega$
system of states, forming $\ell$+1 degenerate levels,
by introducing the pairs
\begin{equation}\label{eqlp1}
A^i = \sum_{m=1}^\Omega a^+_{i m} a^+_{0 m} ,
\end{equation}
where the index 0 refers to one of the levels,
and $i=1,\ldots,\ell$.}
The simplest example is provided by the well-known quasispin
SU(2) algebra
obtained for $\ell$=1.
By normalizing the collective pairs so that $g$=$\Omega$, one obtains
$\Omega$-independent structure constants:
   \begin{equation}\label{e138}
   c^{jl}_{ik} = \delta^j_i\delta^l_k + \delta^j_k\delta^l_i .
   \end{equation}

The mapping of this algebra derived in Ref.\cite{DSG93} from
supercoherent states is
\begin{mathletters}\begin{eqnarray}
   A^j               &\longleftrightarrow&
                       -B^j N_{\rm B}
                     + B^i \big[{\cal A}_i,{\cal A}^j\big]
                        + {\cal A}^j    , \label{su140c} \\
     A_j                    &\longleftrightarrow&
                       B_j                    , \label{su140a} \\
   \big[A_i,A^j\big] &\longleftrightarrow&
                       \delta^j_i(\Omega-N_{\rm B}) - B^jB_i
                                                 \nonumber \\ &&
                     - \left(\Omega\delta^j_i
                 - \big[{\cal A}_i,{\cal A}^j\big]\right)
                                              , \label{su140b} \\
   a^{\nu}           &\longleftrightarrow&
                       \alpha^{\nu}
                 + B^i \big[{\cal A}_i,\alpha^{\nu}\big]
                                              , \label{su140e} \\
   a_{\nu}           &\longleftrightarrow&
                       \alpha_{\nu}
                                              , \label{su140d}
\end{eqnarray}\end{mathletters}%
where $N_{\rm B}$=$B^kB_k$ is the boson-number operator.

Again we may express the similarity transformation in a compact
form as the invariant operator of the fermion core algebra depends
only on the ideal fermion number operator $n=\alpha^\rho \alpha_\rho$,
namely,
\begin{equation}
C_{\rm F} =  \case{1}{2}n(\Omega+1-\case{1}{2}n) \:\:.
\end{equation}
We then get
\begin{equation}\label{eqsu1}
C_{\rm F} - \widehat{C}_{\rm F} =
\case{1}{2}(n-\hat{n})(\Omega+1-\case{1}{2}(n+\hat{n})) ,
\end{equation}
and the transformation (\ref{eq3}) is
\begin{equation}\label{eqsu2}
 X = \frac{(\Omega-\case{1}{2}(n+\hat{n}))!}{(\Omega-\hat{n})!}
     {\rm exp}\left[{\cal A}^i B_i \right] \:
{\textstyle \raisebox{-1ex}{$\widehat{}$}}\qquad.
\end{equation}
The transformed Dyson boson-fermion mapping is obtained in the form
\widetext
\begin{mathletters}\label{sulmap}
\begin{eqnarray}
A^j &\longleftrightarrow&
                       -B^j N_{\rm B}
                     + B^i \big[{\cal A}_i,{\cal A}^j\big]
                                              , \label{su141c} \\
A_j  &\longleftrightarrow&
                       B_j                    , \label{su141a} \\
\big[A_i,A^j\big] &\longleftrightarrow&
                       \delta^j_i(\Omega-N_{\rm B}) - B^jB_i
                     - \left(\Omega\delta^j_i
                 - \big[{\cal A}_i,{\cal A}^j\big]\right)
                                              , \label{su141b} \\
a^{\nu} &\longleftrightarrow&
                       \alpha^{\nu}
                 + B^i \big[{\cal A}_i,\alpha^{\nu}\big]
 -\frac{1}{\Omega-n+2} {\cal A}^i \big[{\cal A}_i,\alpha^\nu\big]
+\frac{1}{\Omega-n+1} B^i\big[\big[{\cal A}_i,\alpha^\nu\big]
,{\cal A}^l\big] B_l   \nonumber \\
& & +\frac{1}{(\Omega-n+1)(\Omega-n+2)} {\cal A}^l B^i\big[{\cal A}_i
,\alpha^\nu\big] B_l
                                              , \label{su141e} \\
a_{\nu} &\longleftrightarrow&
                       \alpha_{\nu}
 +\frac{1}{(\Omega-n+1)(\Omega-n+2)} {\cal A}^l B_l \alpha_\nu
 +\frac{1}{\Omega-n+1} \big[\alpha_\nu,{\cal A}^l\big] B_l
                                              . \label{su141d}
\end{eqnarray}\end{mathletters}\narrowtext%

As in the SO(2$N$) case we observe that the single fermion images are
finite and contain terms changing the ideal fermion number by one
only.  For the SU(2) algebra the mapping (\ref{sulmap})
reduces to the one derived previously in Ref.\cite{GH83}.
Furthermore, the images (\ref{su141e}) and (\ref{su141d})
preserve anti-commutation relations on the full ideal space, as can be
verified after some algebra.  In
Ref.\cite{GH83} the construction of these same images allowed a simple
argument to prove that these relations were preserved at least on the
physical subspace.  That we could anticipate and finally make a more
general and complete statement about the anti-commutation relatios
here, once again illustrates the versatility of the (super)coherent
state method.

\subsection{SO(5) mapping}
\label{sec5a}

Probably the simplest model with an invariant operator $C_{\rm F}$
which does not
just trivially depend on the fermion number, as in SO(2$N$) and
SU($\ell$+1),
is the SO(5) model\cite{KM91}. It assumes two single-$j$ shells
with the same degeneracy $\Omega$=2$j$+1,
where three kinds of monopole pairs can be formed.
Let us denote these shells by $p$ and $h$
and introduce the following pairs
with the normalization $g$ in (\ref{e60}) chosen to be
equal to $\Omega$:
\widetext
\begin{mathletters}\label{eqso5}\begin{eqnarray}
S_+ &=& \sqrt{\Omega} (a^p a^h)^{(0)} \:\: , \:\:S_- = (S_+)^+ \;\; ,
\:\:S_0 = {\textstyle{1\over 4}} (n_p+n_h-2\Omega) , \label{eqso5a} \\
L_+ &=& \sqrt{\Omega/2} (a^p a^p)^{(0)} \:\: , \:\:L_- = (L_+)^+
\;\; ,
\:\:L_0 = {\textstyle{1\over 2}} (n_p-\Omega) , \label{eqso5b} \\
K_+ &=& \sqrt{\Omega/2} (a^h a^h)^{(0)} \:\: , \:\:K_- = (K_+)^+
\;\; ,
\:\:K_0 = {\textstyle{1\over 2}} (n_h-\Omega) , \label{eqso5c}
\end{eqnarray}\end{mathletters}%
where the fermion creation operators are coupled to angular momentum
zero. The algebra is closed by monopole single-particle operators:
\begin{equation}\label{eqso5d}
T_+ = -\sqrt{\Omega} (a^p \tilde{a}_h)^{(0)} \:\: , \:\:T_- = (T_+)^+
\;\; ,
\:\:T_0 = {\textstyle{1\over 4}} (n_p-n_h) ,
\end{equation}
where the tilde denotes the time-reversed operator.

The boson-fermion mapping derived from
supercoherent state method is (see also Ref.\cite{GH80b} where
a different normalization is used)
\begin{mathletters}\label{eqso5sup}\begin{eqnarray}
S_+ &\longleftrightarrow& B^f(\Omega-{\textstyle{1\over 2}}
     (N_f+n_p+n_h) -N_p-N_h)
    - B^pB^hB_f - B^p {\cal T}_-
   - B^h {\cal T}_+ + {\cal S}_+ ,\label{eqso5supa} \\
S_- &\longleftrightarrow& B_f ,\label{eqso5supb} \\
S_0 &\longleftrightarrow& {\textstyle{1\over 2}} (N_p+N_h +N_f
    +{\textstyle{1\over 2}}n_p +
{\textstyle{1\over 2}}n_h -\Omega) ,\label{eqso5supc} \\
L_+ &\longleftrightarrow& B^p(\Omega-N_p-N_f-n_p)
    -{\textstyle{1\over 2}}B^fB^fB_h - B^f{\cal T}_+
  + {\cal L}_+ ,\label{eqso5supd} \\
L_- &\longleftrightarrow& B_p ,\label{eqso5supe} \\
L_0 &\longleftrightarrow& {\textstyle{1\over 2}} (2 N_p +N_f +n_p -\Omega)
,\label{eqso5supf} \\
K_+ &\longleftrightarrow& B^h(\Omega-N_h-N_f-n_h)
    -{\textstyle{1\over 2}}B^fB^fB_p - B^f{\cal T}_-
  + {\cal K}_+ ,\label{eqso5supg} \\
K_- &\longleftrightarrow& B_h ,\label{eqso5suph} \\
K_0 &\longleftrightarrow& {\textstyle{1\over 2}} (2 N_h +N_f +n_h -\Omega)
,\label{eqso5supi} \\
T_+ &\longleftrightarrow& B^pB_f+B^fB_h
    +{\cal T}_+ ,\label{eqso5supj} \\
T_- &\longleftrightarrow& B^fB_p+B^hB_f
    +{\cal T}_- ,\label{eqso5supk} \\
T_0 &\longleftrightarrow& {\textstyle{1\over 2}}
     (N_p+{\textstyle{1\over 2}}n_p- N_h
    -{\textstyle{1\over 2}}n_h)
,\label{eqso5supl} \\
a^p &\longleftrightarrow& \alpha^p+{\textstyle{1\over\sqrt{2}}} B^f
  \tilde{\alpha}_h
+ B^p\tilde{\alpha}_p ,\label{eqso5supm} \\
a_p &\longleftrightarrow& \alpha_p ,\label{eqso5supn} \\
a^h &\longleftrightarrow& \alpha^h+
     {\textstyle{1\over \sqrt{2}}} B^f \tilde{\alpha}_p
     + B^h\tilde{\alpha}_h ,\label{eqso5supo} \\
a_h &\longleftrightarrow& \alpha_h \; .
\label{eqso5supp}
\end{eqnarray}\end{mathletters}\narrowtext\noindent%
The ideal-fermion-pair and the single-particle
operators are defined in the same way as
in (\ref{eqso5}) and (\ref{eqso5d})
with the $a$ operators replaced by $\alpha$.

The similarity transformation is given by
\begin{equation}\label{eqso5sim}
 X = \sum_{k=0}^{\infty} (\frac{1}{C_{\rm F} - \widehat{C}_{\rm F}} W)^{k}\:
{\textstyle \raisebox{-1ex}{$\widehat{}$}}\qquad,
\end{equation}
with
\begin{equation}\label{eqso5sima}
W = {\cal S}_+ B_f + {\cal L}_+ B_p + {\cal K}_+ B_h
\end{equation}
and
 \begin{equation}\label{eqso5simb}
C_{\rm F} = {\cal S}_+ {\cal S}_- + {\cal L}_+ {\cal L}_-
                                  + {\cal K}_+ {\cal K}_-
\;\; .
\end{equation}
Consequently, the transformed mapping
of the bifermion operators is identical to that given
in Eqs.{\ }(\ref{eqso5supa})---(\ref{eqso5supp}),
except from the fact that the ideal fermion pairs
${\cal S}_+$, ${\cal L}_+$, and ${\cal K}_+$
disappear from
Eqs.{\ }(\ref{eqso5supa}), (\ref{eqso5supd}), and (\ref{eqso5supg}),
respectively.
On the other hand, the mapping of single-fermion operators now reads
\widetext
\begin{mathletters}\label{eqso5tr}\begin{eqnarray}
a^p &\longleftrightarrow& \alpha^p+
      {\textstyle{1\over \sqrt{2}}} B^f \tilde{\alpha}_h
      + B^p\tilde{\alpha}_p
      -\frac{1}{C_{\rm F} - \widehat{C}_{\rm F}} W \:
      {\textstyle
      \raisebox{-1ex}{$\widehat{}$}}\:\:({\textstyle{1\over
      \sqrt{2}}} B^f \tilde{\alpha}_h+ B^p\tilde{\alpha}_p) \nonumber
\\
& &+({\textstyle{1\over \sqrt{2}}} B^f \tilde{\alpha}_h+ B^p\tilde{\alpha}_p)
\frac{1}{C_{\rm F} - \widehat{C}_{\rm F}} W \:
{\textstyle \raisebox{-1ex}{$\widehat{}$}}\:\: + \ldots
,\label{eqso5trm} \\
a_p &\longleftrightarrow& \alpha_p
-\frac{1}{C_{\rm F} - \widehat{C}_{\rm F}} W \:
{\textstyle \raisebox{-1ex}{$\widehat{}$}}\:\:\alpha_p
+\alpha_p\frac{1}{C_{\rm F} - \widehat{C}_{\rm F}} W \:
{\textstyle \raisebox{-1ex}{$\widehat{}$}}\:\: + \ldots
,\label{eqso5trn} \\
a^h &\longleftrightarrow& \alpha^h+{\textstyle{1\over \sqrt{2}}} B^f
      \tilde{\alpha}_p + B^h\tilde{\alpha}_h
    -\frac{1}{C_{\rm F} - \widehat{C}_{\rm F}} W \:
{\textstyle \raisebox{-1ex}{$\widehat{}$}}\:\:({\textstyle{1\over \sqrt{2}}}
B^f \tilde{\alpha}_p+ B^h\tilde{\alpha}_h) \nonumber \\
& &+({\textstyle{1\over \sqrt{2}}} B^f \tilde{\alpha}_p+ B^h\tilde{\alpha}_h)
\frac{1}{C_{\rm F} - \widehat{C}_{\rm F}} W \:
{\textstyle \raisebox{-1ex}{$\widehat{}$}}\:\: + \ldots
,\label{eqso5tro} \\
a_h &\longleftrightarrow& \alpha_h
-\frac{1}{C_{\rm F} - \widehat{C}_{\rm F}} W \:
{\textstyle \raisebox{-1ex}{$\widehat{}$}}\:\:\alpha_h
+\alpha_h\frac{1}{C_{\rm F} - \widehat{C}_{\rm F}} W \:
{\textstyle \raisebox{-1ex}{$\widehat{}$}}\:\: + \ldots
\quad.\label{eqso5trp}
\end{eqnarray}\end{mathletters}%

It is not difficult to perform explicit calculations in this model
and present explicit results for some particular configurations.
For example, it is straightforward to show that the image of $a_p$
(\ref{eqso5trn}) acting on the $|n_{\rm F}=0)$ subspace has the form
\begin{equation}\label{eqso5nf0}
a_p |n_{\rm F}=0\rangle\longleftrightarrow \frac{1}{\Omega}
\left( [\alpha_p, {\cal S}_+]
B_f + [\alpha_p,{\cal L}_+] B_p\right) |n_{\rm F}=0) ,
\end{equation}
where $|n_{\rm F}=0\rangle$ means that all the fermions form collective pairs
only. The image of the same operator acting on the
subspace $|n_{\rm F}(p)=1)$ is as follows
\begin{eqnarray}\label{eqso5nf1}
& &a_p |n_{\rm F}(p)=1\rangle\longleftrightarrow \left(\alpha_p
- \frac{1}{\Omega}\left( {\cal S}_+
B_f + {\cal L}_+ B_p + {\cal K}_+ B_h\right)\alpha_p
\right. \nonumber \\
& & + \frac{1}{(\Omega-1)(\Omega+{\textstyle{1\over 2}})}
\left((\Omega \alpha_p
{\cal S}_+ - \alpha_p {\cal L}_+ {\cal T}_-) B_f +
( (\Omega-{\textstyle{1\over 2}})
{\cal K}_+ \alpha_p -\alpha_p {\cal S}_+ {\cal T}_-)B_h\right)
\nonumber \\
& & \left. + \frac{1}{\Omega-1} \alpha_p {\cal L}_+ B_p \right)
|n_{\rm F}(p)=1) , \end{eqnarray}
where $|n_{\rm F}(p)=1\rangle$ means that one fermion is unpaired
and occupies the level $p$. In Appendix \ref{appB} we show
how the general structure of the single-fermion images
specifies to the case of the SO(5) model.

\subsection{SO(8) mapping}
\label{sec5}

The SO(8) model\cite{Gin80} is defined by collective pairs
\begin{mathletters}\begin{eqnarray}
   F^+_{JM} &=&
   \sqrt{\case{1}{2}} \sum_{j_1j_2}(-1)^{J+i+k+j_1}
   \frac{\hat{j_1}\hat{j_2}}{\hat{k}}
   \left\{\begin{array}{ccc} j_1 & j_2 & J \\
                             i   & i     & k     \end{array}\right\}
   \left(a^+_{j_1}a^+_{j_2}\right)^{(J)}_{M} , \label{so81} \\
   P_{JM} &=&
   -\sqrt{2\Omega} \sum_{j_1j_2}(-1)^{J+i+k+j_1}
   \frac{\hat{j_1}\hat{j_2}}{\hat{k}}
   \left\{\begin{array}{ccc} j_1 & j_2 & J \\
                             i   & i     & k     \end{array}\right\}
   \left(a^+_{j_1}\tilde{a}_{j_2}\right)^{(J)}_{M} , \label{so82}
   \end{eqnarray}\end{mathletters}%
with $i=\frac{3}{2}$ and $k$ integer. In (\ref{so81}) only
$S^+$ ($J$=0) and $D^+$ ($J$=2)
pairs are allowed, while in (\ref{so82}) $J$ takes values
0, 1, 2, and 3.

In order to generalize the model to odd systems we may
add creation and annihilation operators,
$a^\dagger_{jm}$ and $a_{jm}$.
The boson-fermion mapping of this algebra derived from
supercoherent states is
   \begin{mathletters}\begin{eqnarray}
   F^+_{JM} &\longleftrightarrow&  B^{\dagger}_{JM}
      -\frac{2}{\hat{k}^2} \sum_{J_1J_2J_3J'}
       \hat{J_1}\hat{J_2}\hat{J_3}\hat{J'}
          \left\{ \begin{array}{ccc} {i}   & {i}   & J   \\
                                     {i}   & {i}   & J_3 \\
                                     J_2 & J_1 & J'  \end{array}\right\}
             ((B^{\dagger}_{J_1} B^{\dagger}_{J_2})^{(J')}
                   \tilde{B}_{J_3})^{(J)}_{M}  \nonumber \\
   & & + \frac{2}{\hat{i}\hat{k}^2}(-1)^{J_2}\hat{J}_1\hat{J}_2
   \left\{ \begin{array}{ccc} J_1 & J_2 & J \\
                                {i}  &  {i}  & {i} \end{array}\right\}
   \left(B^\dagger_{J_1} {\cal P}_{J_2}\right)^{(J)}_M
   + {\cal F}^\dagger_{JM}  , \label{eqso811}\\
   F_{JM} &\longleftrightarrow& B_{JM} , \label{eqso812}\\
   P_{JM} &\longleftrightarrow& \frac{2\sqrt{2\Omega}}
            {\hat{{k}}}(-1)^{J+2{i}}
            \sum_{J_1J_2}\hat{J}_1 \hat{J}_2
            \left\{ \begin{array}{ccc} J_1 & J_2 & J \\
                               {i}  &  {i}  & {i} \end{array}\right\}
   (B^{\dagger}_{J_1} \tilde{B}_{J_2})^{(J)}_{M} + {\cal P}_{JM} ,
  \label{eqso813}\\
  a^+_{jm} &\longleftrightarrow& \alpha^\dagger_{jm} +
\frac{\sqrt{2}}{\hat{k}}
   \hat{J}_1 \hat{j}_1 (-1)^{j_1+i+k}
\left\{ \begin{array}{ccc} j_1 & j & J_1 \\
       {i}  &  {i}  & {k} \end{array}\right\}
  \left(B^\dagger_{J_1}\tilde{\alpha}_{j_1}\right)^{(j)}_m
  , \label{eqso814}\\
  a_{jm} &\longleftrightarrow& \alpha_{jm} , \label{eqso815}
   \end{eqnarray}\end{mathletters}%
where $\tilde{B}_{JM}$=$(-1)^{J-M}B_{J,-M}$ and
$\tilde{a}_{jm}$=$(-1)^{j-m}a_{j,-m}$.
The ideal fermion pair operators
${\cal F}^\dagger_{JM}$ and ${\cal P}_{JM}$ are given by
Eqs.{\ }(\ref{so81}) and
(\ref{so82}), respectively, with the fermion operators $a_{jm}$ replaced by
ideal fermion operators $\alpha_{jm}$.

In this case we cannot simplify the similarity transformation
\begin{equation}\label{eqso86}
 X = \sum_{k=0}^{\infty} (\frac{1}{C_{\rm F} - \widehat{C}_{\rm F}}
{\cal F}^\dagger_J \cdot \tilde{B}_J)^{k}\:
{\textstyle \raisebox{-1ex}{$\widehat{}$}}\qquad,
\end{equation}
as
\begin{equation}\label{eqso85}
C_{\rm F} - \widehat{C}_{\rm F} = \frac{1}{\Omega}\left[\case{1}{2}(n-\hat{n})
(\Omega+6-\case{1}{2}(n+\hat{n})) +
\hat{C}_{2 {\rm Spin_{\rm F}(6)}}-C_{2 {\rm Spin_{\rm F}(6)}}\right]
,
\end{equation}
with $n$ the ideal fermion number operator and $C_{2 {\rm Spin_{\rm F}(6)}}=
{\textstyle{1\over 4}}(P_1\cdot P_1+P_2\cdot P_2+P_3\cdot P_3)$.
The transformed mapping then reads
   \begin{mathletters}\label{so8images}
\begin{eqnarray}
   F^+_{JM} &\longleftrightarrow& B^{\dagger}_{JM}
      -\frac{2}{\hat{k}^2} \sum_{J_1J_2J_3J'}
       \hat{J_1}\hat{J_2}\hat{J_3}\hat{J'}
          \left\{ \begin{array}{ccc} {i}   & {i}   & J   \\
                                     {i}   & {i}   & J_3 \\
                                     J_2 & J_1 & J'  \end{array}\right\}
             ((B^{\dagger}_{J_1} B^{\dagger}_{J_2})^{(J')}
                   \tilde{B}_{J_3})^{(J)}_{M}  \nonumber \\
   & & + \frac{2}{\hat{i}\hat{k}^2}(-1)^{J_2}\hat{J}_1\hat{J}_2
   \left\{ \begin{array}{ccc} J_1 & J_2 & J \\
                        {i}  &  {i}  & {i} \end{array}\right\}
   \left(B^\dagger_{J_1} {\cal P}_{J_2}\right)^{(J)}_M , \\
   F_{JM} &\longleftrightarrow& B_{JM} , \\
   P_{JM} &\longleftrightarrow& \frac{2\sqrt{2\Omega}}
            {\hat{{k}}}(-1)^{J+2{i}}
            \sum_{J_1J_2}\hat{J}_1 \hat{J}_2
            \left\{ \begin{array}{ccc} J_1 & J_2 & J \\
                               {i}  &  {i}  & {i} \end{array}\right\}
   (B^{\dagger}_{J_1} \tilde{B}_{J_2})^{(J)}_{M} + {\cal P}_{JM}
,\label{eqso87}
\end{eqnarray}\end{mathletters}%
for the collective pair operators while for the single-fermion operators
we obtain
\begin{mathletters}\begin{eqnarray}
a^+_{jm} &\longleftrightarrow& \alpha^\dagger_{jm}
+ B^\dagger_J \cdot \left[\tilde{{\cal F}}_J,\alpha^\dagger_{jm}\right]
           - \frac{1}{C_{\rm F} - \widehat{C}_{\rm F}}
{\cal F}^\dagger_{J_1}\cdot \tilde{B}_{J_1} \:
\:{\textstyle \raisebox{-1ex}{$\widehat{}$}}\:\:
B^\dagger_J \cdot \left[\tilde{{\cal F}}_J,\alpha^\dagger_{jm}\right]
\nonumber  \\
& & + B^\dagger_J \cdot \left[\tilde{{\cal F}}_J,\alpha^\dagger_{jm}\right]
\frac{1}{C_{\rm F} - \widehat{C}_{\rm F}}
{\cal F}^\dagger_{J_1}\cdot \tilde{B}_{J_1} \:
\:{\textstyle \raisebox{-1ex}{$\widehat{}$}}\:\:\ldots\:\:\:
                                              , \label{so812} \\
  a_{jm} &\longleftrightarrow& \alpha_{jm}
- \frac{1}{C_{\rm F} - \widehat{C}_{\rm F}}
{\cal F}^\dagger_{J_1}\cdot \tilde{B}_{J_1} \:
{\textstyle \raisebox{-1ex}{$\widehat{}$}}\: \: \alpha_{jm}
 + \alpha_{jm} \frac{1}{C_{\rm F} - \widehat{C}_{\rm F}}
{\cal F}^\dagger_{J_1}\cdot \tilde{B}_{J_1} \:
{\textstyle \raisebox{-1ex}{$\widehat{}$}}\:\:\ldots\:\:\:
                                              . \label{so826}
\end{eqnarray}\end{mathletters}%
The commutator
$B^\dagger_{J_1} \cdot \left[\tilde{{\cal
F}}_{J_1},\alpha^\dagger_{jm}\right]$
in (\ref{so812}) just gives the second term in (\ref{eqso814}) and
the dots $\dots$ refer to the same class of terms as in
Eqs.\,(\ref{s126e}) and (\ref{s126d}).

Let us now consider the matrix elements of the mapped single-fermion
operators (\ref{so812}) and (\ref{so826}) between even and odd
states $|\Psi\rangle$ and $|\Psi'\rangle$
corresponding to nuclei with particle number differing
by one. It is assumed that the even state contains collective
pairs only.
These matrix elements give then the spectroscopic
factors and can be written in the following form
\begin{mathletters}\label{SO830}\begin{eqnarray}
\langle\Psi|a^+_{jm}|\Psi'\rangle &=&
(\Psi|\Bigl[\alpha^\dagger_{jm}
+\frac{\sqrt{2}}{\hat{k}}
   \hat{J}_1 \hat{j}_1 (-1)^{j_1+i+k}
\left\{ \begin{array}{ccc} j_1 & j & J_1 \\
       {i}  &  {i}  & {k} \end{array}\right\}
\left(B^\dagger_{J_1}\tilde{\alpha}_{j_1}\right)^{(j)}_m \nonumber \\
& & +\frac{2}{\hat{k}^2} \sum_{J_1j_2J_3j'}
       \hat{J_1}\hat{j_2}\hat{J_3}\hat{j'}
          \left\{ \begin{array}{ccc} {k}   & {i}    & j_2 \\
                                     {i}   & {i}   & J_1 \\
                                     j & J_3 & j'  \end{array}\right\}
             ((B^{\dagger}_{J_1} \alpha^{\dagger}_{j_2})^{(j')}
                   \tilde{B}_{J_3})^{(j)}_{M}
                        \Big]|\Psi')
                                              , \label{so831} \\
  \langle\Psi|\tilde{a}_{jm}|\Psi'\rangle &=&
   (\Psi|\Big[\tilde{\alpha}_{jm}
-\frac{\sqrt{2}}{\hat{k}}
   \hat{J}_1 \hat{j}_1 (-1)^{j+i+k}
\left\{ \begin{array}{ccc} j_1 & j & J_1 \\
       {i}  &  {i}  & {k} \end{array}\right\}
\left(\alpha^\dagger_{j_1}\tilde{B}_{J_1}\right)^{(j)}_m
                        \Big]|\Psi')
                                              . \label{so832}
\end{eqnarray}\end{mathletters}\narrowtext%

Note that unlike the phenomenological interacting boson-fermion
model (IBFM) case\cite{BBI81} the
tensorial form of the appropriate single-fermion transfer operators,
as functions of supergenerators and ideal fermions [compare
Eqs.{} (\ref{eqsso2n}]
or their collective counterparts), is here {\em
uniquely} fixed by the mapping.  For example in the U(6/4) case
[$k$=0
in Eqs.{\ }(\ref{SO830})] the operators have
the $\sigma_1=\case{1}{2}$
Spin(6) tensorial character.

It should be mentioned that a boson-fermion analysis of the
SO(8) model had previously been presented by Frank {\it et al}.{}
\cite{FHCP87} from the point of view of group contractions. Although
this led to the emergence of an IBFM-type structure,  only
some truncated Holstein-Primakoff images of the SO(8)
generators were
presented, which makes direct comparison with our exact Dyson images
(\ref{so8images}) difficult. Furthermore,
the construction of single
fermion images is not considered at all in Ref.{}\cite{FHCP87}, while
the identification of possible supersymmetric structures is only
speculated about. We refer to Ref.{}\cite{NGD94} for further
discussion concerning supersymmetry in this context, as well as for a
concrete example.

Finally, it is important to realize that the operator images of
$a^+_{jm}$ and $\tilde{a}_{jm}$ which are effective in
Eqs.{\ }(\ref{SO830}) cannot simply be compounded to obtain the image
of an interaction term of the type $a^+a^+aa$, say, as these images
are valid only for the subspace with zero and one ideal fermions only.
This type of inconsistent application in some semi-microscopic
applications is also analyzed and discussed in Ref.\cite{GM89}.

\section{Conclusions}
\label{sec7}

In this paper we have derived a generalized Dyson boson-fermion
mapping of the most general collective fermion-pair algebra extended
by single-fermion operators.  The mapping is given as finite
non-hermitian boson-fermion images of fermion pairs and single-fermion
operators, both expressed in terms of ideal boson and ideal fermion
annihilation and creation operators.

The constructed mapping exploits an important freedom
available in the boson-fermion space, namely that suitable similarity
transformations can be devised to shift between components of
the boson sector and the ideal-fermion-pair sector of the ideal space.
The principal achievement of the present paper lies in finding
an explicit form of such a similarity transformation which leads
to collective even states being mapped onto boson states only, while
collective odd
states are mapped onto boson-fermion states with one ideal fermion
only. In algebraic models we are thus able to account fully
for the effects of the Pauli correlations between the odd
fermion and the collective fermion pairs.

Although our results are based on stringent conditions of algebra
closure, they may serve as guidelines for realistic cases where the
exact closure need not be fulfilled.  In particular, by using our
boson-fermion mapping we have been able to derive microscopically some
supersymmetric structures\cite{NGD94} which previously have only been
introduced in a phenomenological way.  Similarly, we can explicitly
obtain the tensorial structure of the single-fermion operators
defining the spectroscopic factors in boson-fermion models.

We have derived a general formula for the required similarity
transformation as a power series in terms of a particular operator
invariant with respect to the ideal-fermion core subalgebra.  In two
cases, for the SO(2$N$) and SU($\ell$+1) models, we can sum the series
and give finite expressions for the images of single-fermion
operators.  We also showed that in the general case simple physical
requirements allow the series to be truncated to low-order terms when
one is e.g.{\ }interested in calculating spectroscopic factors.

Our two-stage construction of boson-fermion images, viz.{\ }a first
image deduced from a supercoherent state, followed by a similarity
transformation, allowed us to demonstrate that the single fermion
images constructed in this way generally preserve anti-commutation
relations on the full ideal space, unlike previous constructions of
many workers in this field where these relations were valid on the
physical subspace only.  We anticipate that our more general result
will become important especially when boson-fermion calculations are
carried out in the full ideal space as part of the program discussed
and advocated in Ref.{}\cite{DGH91}.

\acknowledgements{This work was supported by grants from the Foundation for
Research Development of South Africa, the University of
Stellenbosch, and in part by the Polish State Committee for
Scientific Research under Contract No.~2~P03B~034~08 and through
the Czech Republic grants GA CR No. 202/93/2472 and GA ASCR
A1048504.}

\appendix
\widetext
\section{Structure of single-fermion images in the general case}
\label{appA}

We consider here the second-order terms in the annihilation operator
(\ref{eqx4}) containing two  boson operators $B$,
\begin{eqnarray}
& &\left( \alpha_\nu \frac{1}{C_{\rm F} - \widehat{C}_{\rm F}}{\cal A}^l B_l \:
    \frac{1}{C_{\rm F} - \widehat{C}_{\rm F}}{\cal A}^m B_m \:
\:{\textstyle \raisebox{-1ex}{$\widehat{}$}}\:\:
-\frac{1}{C_{\rm F} - \widehat{C}_{\rm F}}{\cal A}^l B_l \:
    \frac{1}{C_{\rm F} - \widehat{C}_{\rm F}}{\cal A}^m B_m \:
\:{\textstyle \raisebox{-1ex}{$\widehat{}$}}\:\: \alpha_\nu
\right.                                      \nonumber \\
&& -\Bigl(\frac{1}{C_{\rm F} - \widehat{C}_{\rm F}}{\cal A}^l B_l \:
    \:{\textstyle \raisebox{-1ex}{$\widehat{}$}}\:\:\Bigr) \alpha_\nu
    \Bigl(\frac{1}{C_{\rm F} - \widehat{C}_{\rm F}}{\cal A}^m B_m \:
    \:{\textstyle \raisebox{-1ex}{$\widehat{}$}}\:\:\Bigr)
\nonumber \\
&&\left. +\Bigl(\frac{1}{C_{\rm F} - \widehat{C}_{\rm F}}{\cal A}^l B_l \:
    \:{\textstyle \raisebox{-1ex}{$\widehat{}$}}\:\:\Bigr)
    \Bigl(\frac{1}{C_{\rm F} - \widehat{C}_{\rm F}}{\cal A}^m B_m \:
    \:{\textstyle \raisebox{-1ex}{$\widehat{}$}}\:\:\Bigr)
    \alpha_\nu\right) |\psi)
\nonumber \\
&=&\left( \alpha_\nu \frac{1}{C_{\rm F}}{\cal A}^l B_l \:
    \frac{1}{C_{\rm F}}{\cal A}^m B_m \:
-\frac{1}{C_{\rm F}}{\cal A}^l B_l \:
    \frac{1}{C_{\rm F}}{\cal A}^m B_m \:\: \alpha_\nu
\right.                                      \nonumber \\
&&- \left.\frac{1}{C_{\rm F} - \widehat{C}_{\rm F}}{\cal A}^l B_l \:
    \:{\textstyle \raisebox{-1ex}{$\widehat{}$}}\:\: \alpha_\nu
    \frac{1}{C_{\rm F}}{\cal A}^m B_m \:
+\frac{1}{C_{\rm F} - \widehat{C}_{\rm F}}{\cal A}^l B_l \:
    \:{\textstyle \raisebox{-1ex}{$\widehat{}$}}\:\:
    \frac{1}{C_{\rm F}}{\cal A}^m B_m \:\: \alpha_\nu\right) |\psi) \:\:,
\label{eqn21}
\end{eqnarray}
where (\ref{eqn20}) is assumed. We furthermore investigate
\begin{equation}\label{eqn22}
C_{\rm F} \alpha_\nu \frac{1}{C_{\rm F}}{\cal A}^l B_l \:
    \frac{1}{C_{\rm F}}{\cal A}^m B_m \: |\psi)
      = {\cal A}^i \alpha_\nu{\cal A}_i
    \frac{1}{C_{\rm F}}{\cal A}^l B_l \:\frac{1}{C_{\rm F}}{\cal A}^m B_m \:
    |\psi)\; .
\end{equation}
To evaluate
\begin{equation}\label{eqn24}
{\cal A}_i\frac{1}{C_{\rm F}}{\cal A}^l B_l
                  \:\frac{1}{C_{\rm F}}{\cal A}^m B_m \: |\psi)
\end{equation}
we perform an auxiliary bosonization of the
fermion operators ${\cal A}$  in the spirit of mapping
(\ref{eqsecbos}).
This time, however, we introduce, in addition to the auxiliary bosons
$b$,
also auxiliary fermions $\bar{\alpha}$ commuting with the bosons $B$
and $b$
and consider the mapping (\ref{eqsecbos}) in the alternative
form
   \begin{mathletters}\label{eqsecbosfer}\begin{eqnarray}
   {\cal A}^j               &\longleftrightarrow&  [\bar\Lambda , b^j]
                                               , \label{eq28c} \\
   {\cal A}_j  &\longleftrightarrow&
                                        b_j     , \label{eq28a} \\
   \big[{\cal A}_i,{\cal A}^j\big] &\longleftrightarrow&
                         g\delta^j_i - c^{jl}_{ik}b^kb_l
                         - \chi^j_{\mu\rho}\chi_i^{\nu\rho}
                     \bar{\alpha}^{\mu}\bar{\alpha}_{\nu}
                                              , \label{eq28b} \\
   \alpha^{\nu}    &\longleftrightarrow&
                         \bar{\alpha}^{\nu}
                   + \chi_i^{\nu\rho} b^i\bar{\alpha}_{\rho}
- \frac{1}{C_f - \widehat{C}_f} {\bar{\cal A}}^l b_l\:
\:{\textstyle \raisebox{-1ex}{$\widehat{}$}}\:\:
\chi_n^{\nu\rho}b^n \bar{\alpha}_\rho
                                        \nonumber \\
&+& \chi_n^{\nu\rho}b^n \bar{\alpha}_\rho \frac{1}{C_f - \widehat{C}_f}
{\bar{\cal A}}^l b_l\:
\:{\textstyle \raisebox{-1ex}{$\widehat{}$}}\:\:
\ldots
                                              , \label{eq28e} \\
\alpha_{\nu}      &\longleftrightarrow&
                         \bar{\alpha}_{\nu}
- \frac{1}{C_f - \widehat{C}_f} {\bar{\cal A}}^l b_l\:
{\textstyle \raisebox{-1ex}{$\widehat{}$}}\: \: \bar{\alpha}_\nu
+ \bar{\alpha}_\nu \frac{1}{C_f - \widehat{C}_f} {\bar{\cal A}}^l b_l\:
{\textstyle \raisebox{-1ex}{$\widehat{}$}}\:\:
\ldots
                    , \label{eq28d}
   \end{eqnarray}\end{mathletters}%
with ${\bar{\cal
A}}^l={\textstyle{1\over 2}}\chi_{\mu\nu}^l\bar{\alpha}^\mu\bar{\alpha}^\nu$,
$C_f = {\bar{\cal A}}^l {\bar{\cal A}}_l$, and
\begin{eqnarray}\label{eqn12}
\bar\Lambda &=& g b^l b_l -{\textstyle{1\over 4}}
                c_{m n}^{k l} b^m b^n b_l b_k
           - \chi^k_{\mu\rho}\chi^{\nu\rho}_l b^l b_k\bar{\alpha}^\mu
        \bar{\alpha}_\nu \nonumber \\
        &=& (C_{\rm F})_{\rm bf} + {\textstyle{1\over 4}}
            c_{m n}^{k l} b^m b^n b_l b_k
= {\textstyle{1\over 2}}(C_{\rm F})_{\rm bf} +
  {\textstyle{1\over 2}} b^l [{\bar{\cal A}}_l,{\bar{\cal
   A}}^k] b_k \: ,
\end{eqnarray}
where $(C_{\rm F})_{\rm bf}$ is expressed using the operators $b$ and
$\bar{\alpha}$.
The condition (\ref{eqn20}) in the $b$, $\bar{\alpha}$ space takes the
form $b_l |\psi)_{\rm bf} =0$, that is there
are no bosons $b$ contained in
$|\psi)_{\rm bf}$.

It follows that (\ref{eqn24}) is mapped onto
\begin{eqnarray}\label{eqn13}
& &b_i\frac{1}{(C_{\rm F})_{\rm bf}
- (\widehat{C}_{\rm F})_{\rm bf}}[\bar\Lambda,b^l] B_l\:
\frac{1}{(C_{\rm F})_{\rm bf}
- (\widehat{C}_{\rm F})_{\rm bf}}[\bar\Lambda,b^k] B_k\:
{\textstyle \raisebox{-1ex}{$\widehat{}$}}\: \: |n_b=0) \nonumber \\
&=&b_i{\textstyle{1\over 2}} b^l B_l b^k B_k |n_b=0) = B_i b^k B_k |n_b=0)\:\:
{}.
\end{eqnarray}
This is the image of the original ideal space state
$B_i\frac{1}{C_{\rm F}} {\cal A}^k B_k |\psi)$, which follows after
operating with $\bar\Lambda{\bar\Lambda}^{-1}$ on the final state
above and completing the commutator $[\bar\Lambda,b^k]$ to identify
the image of ${\cal A}^k$. Consequently,
we find from (\ref{eqn22}) the relation
\begin{equation}\label{eqn25}
\alpha_\nu \frac{1}{C_{\rm F}}{\cal A}^l B_l \:
    \frac{1}{C_{\rm F}}{\cal A}^m B_m \: |\psi) = \frac{1}{C_{\rm F}}
{\cal A}^i B_i \alpha_\nu\:\frac{1}{C_{\rm F}}{\cal A}^m B_m \: |\psi) \; .
\end{equation}
We note that it is not possible to repeat a similar derivation
for the first
order terms as $C_{\rm F}$ acting on
$\alpha_\nu \frac{1}{C_{\rm F}}{\cal A}^l B_l |\psi)$
has in general zero eigenvalues -- this is evident for
$|\psi)\equiv|n_{\rm F}=0)$, for example.

Returning to Eq.{\ }(\ref{eqn21})  and using (\ref{eqn25}),
we have after some manipulation
\begin{eqnarray}\label{eqn26}
& &\left(\frac{C_{\rm F} - \widehat{C}_{\rm F}}{C_{\rm F} -
\widehat{C}_{\rm F}}\frac{1}{C_{\rm F}}
{\cal A}^i B_i \:
{\textstyle \raisebox{-1ex}{$\widehat{}$}}\: \:
\alpha_\nu\:\frac{1}{C_{\rm F}}{\cal A}^m B_m \:
- \frac{1}{C_{\rm F} - \widehat{C}_{\rm F}}{\cal A}^l B_l \:
    \:{\textstyle \raisebox{-1ex}{$\widehat{}$}}\:\: \alpha_\nu
    \frac{1}{C_{\rm F}}{\cal A}^m B_m \: \right.
\nonumber \\
&+&\left.\frac{1}{C_{\rm F} (C_{\rm F}-\widehat{C}_{\rm F})}{\cal A}^i B_i \:
{\textstyle \raisebox{-1ex}{$\widehat{}$}}\: \:
{\cal A}^m B_m \:\alpha_\nu\right)|\psi)
=\left( -\frac{1}{C_{\rm F} (C_{\rm F}-\widehat{C}_{\rm F})}{\cal A}^i B_i \:
{\textstyle \raisebox{-1ex}{$\widehat{}$}}\: \:
{\cal A}^k\:\alpha_\nu\:{\cal A}_k\:\frac{1}{C_{\rm F}}{\cal A}^m B_m
\right. \nonumber \\
&+& \left.\frac{1}{C_{\rm F} (C_{\rm F}-\widehat{C}_{\rm F})}{\cal A}^i B_i \:
{\textstyle \raisebox{-1ex}{$\widehat{}$}}\: \:
{\cal A}^m B_m \:\alpha_\nu\right)|\psi) = 0 ,
\end{eqnarray}
because, using (\ref{eqsecbosfer}) again, we have
 ${\cal A}_k \frac{1}{C_{\rm F}}{\cal A}^m B_m |\psi) = B_k |\psi)$.

It is now straightforward to prove the cancellation for all
higher-order terms by induction using
\begin{eqnarray}\label{eqn16}
& &b_l \frac{1}{(C_{\rm F})_{\rm bf}
- (\widehat{C}_{\rm F})_{\rm bf}}[\bar\Lambda,b^{i_1}]
B_{i_1}\:
\ldots
\frac{1}{(C_{\rm F})_{\rm bf}
- (\widehat{C}_{\rm F})_{\rm bf}}[\bar\Lambda,b^{i_n}] B_{i_n}\:
{\textstyle \raisebox{-1ex}{$\widehat{}$}}\: \: |n_b=0)
\nonumber \\
&=& b_l \frac{1}{n!} b^{i_1} B_{i_1}\ldots b^{i_n} B_{i_n} |n_b=0)
= B_l \frac{1}{(n-1)!} b^{i_1} B_{i_1}\ldots b^{i_{n-1}} B_{i_{n-1}} |n_b=0)
\;\;\; .
\end{eqnarray}

Since annihilation and creation operators are linked by the commutation
relations (\ref{e5x}) it follows that the terms increasing the number
of ideal fermions
$\alpha$ by three, five etc.{\ }also cancel in the creation
operator (\ref{eqx5})
when acting on the class of states characterized by
(\ref{eqn20}).

Nevertheless, it is instructive to see explicitly how the terms
increasing the ideal fermion
number by three in the creation operator image actually cancel.
These terms appear
in expression (\ref{eqx5}) which shows that the relevant boson
operator parts have respectively the structure
$B^k B_l B_i$ and that of a single-boson annihilation operator.
The terms of the former type cancel as the
structure is of exactly the same form as the
second-order part of the annihilation operator image for which we have
demonstrated the cancellation above. The latter part can be written as
\begin{eqnarray}\label{eqn27}
 & &\left(\alpha^\nu \frac{1}{C_{\rm F}}{\cal A}^l B_l \:
- \frac{1}{C_{\rm F} - \widehat{C}_{\rm F}}{\cal A}^l B_l \:
\:{\textstyle \raisebox{-1ex}{$\widehat{}$}}\:\: \alpha^\nu
-\frac{1}{C_{\rm F} - \widehat{C}_{\rm F}}{\cal A}^l    \:
    \:{\textstyle \raisebox{-1ex}{$\widehat{}$}}\:\:
    [{\cal A}_l, \alpha^\nu]
    \frac{1}{C_{\rm F}}{\cal A}^m B_m \: \right.
                                         \nonumber \\
 &&+\left.\frac{1}{C_{\rm F} (C_{\rm F}
- \widehat{C}_{\rm F})} {\cal A}^m B_m \:
    \:{\textstyle \raisebox{-1ex}{$\widehat{}$}}\:\:
    {\cal A}^l [{\cal A}_l, \alpha^\nu]
    + \frac{1}{C_{\rm F} (C_{\rm F} - \widehat{C}_{\rm F})} {\cal A}^l \:
    \:{\textstyle \raisebox{-1ex}{$\widehat{}$}}\:\:
    {\cal A}^m B_m [{\cal A}_l, \alpha^\nu] \right) |\psi)
                                      \nonumber \\
&=& \left(\frac{1}{C_{\rm F}} {\cal A}^l
[{\cal A}_l,\alpha^\nu]\frac{1}{C_{\rm F}}
{\cal A}^m B_m
-\frac{1}{C_{\rm F}(C_{\rm F}-\widehat{C}_{\rm F})} {\cal A}^m B_m \:
    \:{\textstyle \raisebox{-1ex}{$\widehat{}$}}\:\:{\cal A}^k {\cal A}_k
    \alpha^\nu
         \right.                               \nonumber \\
 &&- \frac{1}{C_{\rm F} - \widehat{C}_{\rm F}}{\cal A}^l    \:
    \:{\textstyle \raisebox{-1ex}{$\widehat{}$}}\:\:
    [{\cal A}_l, \alpha^\nu]
    \frac{1}{C_{\rm F}}{\cal A}^m B_m \:
                                         \nonumber \\
  &&+\left.\frac{1}{C_{\rm F} (C_{\rm F} -
\widehat{C}_{\rm F})} {\cal A}^m B_m \:
    \:{\textstyle \raisebox{-1ex}{$\widehat{}$}}\:\:
    {\cal A}^l [{\cal A}_l, \alpha^\nu]
    + \frac{1}{C_{\rm F} (C_{\rm F} - \widehat{C}_{\rm F})} {\cal A}^l \:
    \:{\textstyle \raisebox{-1ex}{$\widehat{}$}}\:\:
    {\cal A}^m B_m [{\cal A}_l, \alpha^\nu] \right) |\psi)
                                      \nonumber \\
&=& \left(\frac{C_{\rm F}-\widehat{C}_{\rm F}}{C_{\rm F}
-\widehat{C}_{\rm F}}\frac{1}{C_{\rm F}} {\cal
A}^l\:
    \:{\textstyle \raisebox{-1ex}{$\widehat{}$}}\:\:
[{\cal A}_l,\alpha^\nu]\frac{1}{C_{\rm F}}{\cal A}^m B_m
\right.                                        \nonumber \\
 &&- \left.\frac{1}{C_{\rm F} - \widehat{C}_{\rm F}}{\cal A}^l    \:
    \:{\textstyle \raisebox{-1ex}{$\widehat{}$}}\:\:
    [{\cal A}_l, \alpha^\nu]
    \frac{1}{C_{\rm F}}{\cal A}^m B_m \:
    + \frac{1}{C_{\rm F} (C_{\rm F} - \widehat{C}_{\rm F})} {\cal A}^l \:
    \:{\textstyle \raisebox{-1ex}{$\widehat{}$}}\:\:
    {\cal A}^m B_m [{\cal A}_l, \alpha^\nu] \right) |\psi)
                 \nonumber \\
&=& \left(-\frac{1}{C_{\rm F} (C_{\rm F} - \widehat{C}_{\rm F})} {\cal A}^l \:
    \:{\textstyle \raisebox{-1ex}{$\widehat{}$}}\:\: {\cal A}^k {\cal A}_k
    [{\cal A}_l, \alpha^\nu] \frac{1}{C_{\rm F}}{\cal A}^m B_m
\right. \nonumber \\
    &&+   \left. \frac{1}{C_{\rm F} (C_{\rm F}
- \widehat{C}_{\rm F})} {\cal A}^l \:
    \:{\textstyle \raisebox{-1ex}{$\widehat{}$}}\:\:
    {\cal A}^m B_m [{\cal A}_l, \alpha^\nu] \right) |\psi) = 0\; ,
                                        \label{eqn9}
\end{eqnarray}
using
\begin{equation}\label{eqn10}
\left[ \alpha^\nu , \frac{1}{C_{\rm F}} \right] = \frac{1}{C_{\rm F}}
\left[C_{\rm F} , \alpha^\nu \right] \frac{1}{C_{\rm F}}
\end{equation}
and
\begin{equation}\label{eqn28}
C_{\rm F} [{\cal A}_l, \alpha^\nu] \frac{1}{C_{\rm F}}{\cal A}^m B_m |\psi) =
{\cal A}^m B_m [{\cal A}_l, \alpha^\nu] |\psi)
\end{equation}
derived from expressions (\ref{eqsecbosfer}) and the
definition (\ref{eq3b}).

\section{Structure of single-fermion images in the SO(5) case}
\label{appB}

Here we exemplify
the result of Appendix \ref{appA} in the case of the
SO(5) model discussed in Sec.{\ }\ref{sec5a}. We show
that higher order terms in the single-fermion images cancel for the
second-order part of the annihilation operator acting in the space where no
ideal fermions are present. To be explicit, we show that
\begin{equation}\label{eqso5sec}
\left(\alpha_p \frac{1}{C_{\rm F} - \widehat{C}_{\rm F}} W \:
\frac{1}{C_{\rm F} - \widehat{C}_{\rm F}} W \:
{\textstyle \raisebox{-1ex}{$\widehat{}$}}\:\:
- \Bigl(\frac{1}{C_{\rm F} - \widehat{C}_{\rm F}} W \:
{\textstyle \raisebox{-1ex}{$\widehat{}$}}\:\:\Bigr) \alpha_p \:
\Bigl(\frac{1}{C_{\rm F} - \widehat{C}_{\rm F}} W \:
{\textstyle \raisebox{-1ex}{$\widehat{}$}}\:\:\Bigr)\right)|n_{\rm F}=0) =
0 \; .
\end{equation}
The left hand side follows from (\ref{eqn21}), $W$ is given by
(\ref{eqso5sima}), and $C_{\rm F}$ by (\ref{eqso5simb}).
It can be derived that
\begin{eqnarray}\label{eqso5seca}
& &\frac{1}{C_{\rm F} - \widehat{C}_{\rm F}} W \:\frac{1}{C_{\rm F}
- \widehat{C}_{\rm F}} W \:
{\textstyle \raisebox{-1ex}{$\widehat{}$}}\:\: |n_{\rm F}=0)
\nonumber \\
& = &
\left(\frac{1}{\Omega(\Omega-1)} \Bigl({\cal L}_+{\cal S}_+ B_f B_p
+ {\cal K}_+{\cal S}_+ B_f B_h +
  {\textstyle{1\over 2}}{\cal L}_+{\cal L}_+ B_p B_p
+ {\textstyle{1\over 2}}{\cal K}_+{\cal K}_+ B_h B_h\Bigr)
\right. \nonumber \\
& & +\frac{1}{\Omega(\Omega-1)(2\Omega+1)}\Bigl(\Omega {\cal
S}_+{\cal S}_+ + {\cal K}_+{\cal L}_+\Bigr) B_f B_f
\nonumber \\
& & \left. +\frac{1}{\Omega(\Omega-1)(2\Omega+1)}\Bigl({\cal
S}_+{\cal S}_+
+ (2\Omega-1){\cal K}_+{\cal L}_+ \Bigr) B_p B_h \right) |n_{\rm F}=0)
 \end{eqnarray}
and
\begin{eqnarray}\label{eqso5secb}
& &\Bigl(\frac{1}{C_{\rm F} - \widehat{C}_{\rm F}} W \:
{\textstyle \raisebox{-1ex}{$\widehat{}$}}\:\: \Bigr)\alpha_p \:
\Bigl(\frac{1}{C_{\rm F} - \widehat{C}_{\rm F}} W \:
{\textstyle \raisebox{-1ex}{$\widehat{}$}}\:\:\Bigr)|n_{\rm F}=0)
\nonumber \\
&=&\left(\frac{1}{\Omega(\Omega-1)}\Bigl( ({\cal L}_+[\alpha_p,{\cal
S}_+] +{\cal S}_+[\alpha_p,{\cal L}_+]) B_f B_p
+  {\cal K}_+[\alpha_p,{\cal S}_+ ] B_f B_h
+ {\cal L}_+[\alpha_p,{\cal L}_+] B_p B_p )\Bigr) \right.
\nonumber \\
& &+ \frac{1}{\Omega(\Omega-1)(\Omega+{\textstyle{1\over 2}})}\Bigl(\Omega
{\cal S}_+ [\alpha_p,{\cal S}_+]
+ {\textstyle{1\over 2}}{\cal K}_+[\alpha_p,{\cal L}_+]\Bigr) B_f B_f
\nonumber \\
& & \left.+
\frac{1}{\Omega(\Omega-1)(\Omega+{\textstyle{1\over 2}})}\Bigl({\cal
S}_+ [\alpha_p,{\cal S}_+]
+ (\Omega-{\textstyle{1\over 2}}){\cal K}_+
[\alpha_p,{\cal L}_+] \Bigr) B_p B_h \right) |n_{\rm F}=0) \qquad .
\end{eqnarray}
{}From (\ref{eqso5seca}) and
(\ref{eqso5secb}) it is apparent that (\ref{eqso5sec}) is fulfilled.

It is interesting to note that the above derivation
is valid even for spurious states
in the boson-fermion space. Consider e.g.{\ }$\Omega=1$.
The two-boson configuration
\begin{equation}\label{eqso5spur}
 (B^f B^f + B^p B^h)|0)
\end{equation}
is then a spurious state\cite{DGH91}.
We observe that expressions
(\ref{eqso5nf1}), (\ref{eqso5seca}), and (\ref{eqso5secb})
are singular for $\Omega=1$. However, for $\Omega=1$
these equations are simply not
applicable as they stand, because
\begin{equation}\label{eqso5spur1}
({\cal S}_+{\cal S}_+ + {\cal L}_+{\cal K}_+)|0) = 0
\end{equation}
and
\begin{equation}\label{eqso5spur2}
({\cal S}_+ \alpha^h + {\cal K}_+ \alpha^p) |0) = 0 .
\end{equation}
Consequently, instead of (\ref{eqso5seca}) we have here
\begin{eqnarray}\label{eqso5secc}
& &\frac{1}{C_{\rm F} - \widehat{C}_{\rm F}} W \:\frac{1}{C_{\rm F}
- \widehat{C}_{\rm F}} W \:
{\textstyle \raisebox{-1ex}{$\widehat{}$}}\:\: |n_{\rm F}=0)  \nonumber \\
&= & \frac{4}{3\Omega(2\Omega+1)}({\textstyle{1\over 2}} {\cal
     S}_+{\cal S}_+
+ {\cal K}_+{\cal L}_+) (-{\textstyle{1\over 2}}
  B_f B_f + B_p B_h) |n_{\rm F}=0),
\end{eqnarray}
and (\ref{eqso5secb}) should be replaced by
\begin{eqnarray}\label{eqso5secd}
& &\frac{1}{C_{\rm F} - \widehat{C}_{\rm F}} W \:
{\textstyle \raisebox{-1ex}{$\widehat{}$}}\:\: \alpha_p \:
\frac{1}{C_{\rm F} - \widehat{C}_{\rm F}} W \:
{\textstyle \raisebox{-1ex}{$\widehat{}$}}\:\:|n_{\rm F}=0)  \nonumber \\
&= & \frac{4}{3\Omega(2\Omega+1)}( {\cal S}_+
[\alpha_p,{\cal S}_+]
+ {\cal K}_+[\alpha_p,{\cal L}_+])
(-{\textstyle{1\over 2}} B_f B_f + B_p B_h) |n_{\rm F}=0)
,
\end{eqnarray}
from which we can see that the second-order
contributions
to the single-fermion annihilation operator image cancel even in this
case. Moreover, it is apparent that the operators that act on
$|n_{\rm F}=0)$ in both
 both (\ref{eqso5secc}) and
(\ref{eqso5secd}) give a zero state when they act on the
spurious state (\ref{eqso5spur}).
\narrowtext

\end{document}